\documentclass{article}
 
\newlength{\defbaselineskip}
\RequirePackage[T1]{fontenc}
\RequirePackage[tt=false, type1=true]{libertine}
\RequirePackage[varqu]{zi4}
\RequirePackage[libertine]{newtxmath}
 
\usepackage[authoryear,round,sort&compress]{natbib}

\usepackage[parfill]{parskip}
\usepackage{microtype}
\usepackage{url}
\usepackage{booktabs}
\usepackage{multirow}
\usepackage{tcolorbox}
\usepackage{lineno}

\usepackage[colorlinks=true,linkcolor=blue,citecolor=magenta,urlcolor=red]{hyperref}

\usepackage{graphicx}
\usepackage{arydshln}
\usepackage{xcolor}

\newcommand{\model}[1]{\textsf{#1}}
\newcommand{\rt}[1]{\textbf{\textcolor{red}{#1}}}
\newcommand{\bt}[1]{#1}

\title{Understanding the Effects of Safety Unalignment on Large Language Models}
 
\author{John Halloran \thanks{Alternative contact: halloj3@uw.edu.} \\
Leidos\\
\texttt{halloranjt@leidos.com}
}

\date{}
\begin{document}
 
\maketitle
\begin{abstract}
Safety alignment has become a critical step to ensure LLMs refuse harmful requests while providing helpful and harmless responses.
However, despite the ubiquity of safety alignment for deployed frontier models, 
two separate lines of recent work--\emph{jailbreak-tuning} (JT) and \emph{weight orthogonalization} (WO)--have shown that safety guardrails may be largely disabled, resulting in LLMs which comply with harmful requests they would normally refuse.  In spite of far-reaching safety implications, analysis has largely been limited to refusal rates of each unalignment method in isolation, leaving their relative effects on adversarial LLM capabilities unknown.  To fill this gap, we study the impact of unaligning six popular LLMs of various sizes across a large number of malicious and benign tasks, using both JT and WO.  
Across the evaluated models, we show that while refusal degradation is split between the two methods, WO produces LLMs far more capable of aiding in malicious activity; in contrast to JT, the majority of WO unaligned models are far less prone to hallucinations, better retain their original natural-language performance, and are more effective at state-of-the-art adversarial and cyber attacks.  To thus help mitigate the malicious risks of WO unalignment, we conclude by showing that supervised fine-tuning effectively limits the adversarial attack abilities enabled by WO, without 
drastically affecting hallucination rates or natural language performance.
\end{abstract}

\section{Introduction}

While large language models (LLMs) are powerful productivity-enhancing tools, they also carry serious potential for aiding in malicious activities~\citep{mehrotra2024tree, recordedfuture2024ai_malware}.
Safety alignment (SA) has thus become a critical step to limit such malicious aid, ensuring LLMs refuse harmful requests while providing helpful and harmless responses~\citep{bai2022training, NEURIPS2023_4dbb61cb, dai2024safe, tian2024finetuning, wang2024arithmetic}.  SA consists of multiple rounds of fine-tuning using malicious and benign data.  By leveraging state-of-the-art (SOTA) preference-tuning algorithms~\citep{rafailov2023direct, shao2024deepseekmath, ji2025aialignmentcomprehensivesurvey}, LLMs are trained to refuse malicious requests while following benign instructions, thus equipping models with safety \emph{guardrails}.  
As the catalogue of LLM attacks and malicious uses continues to grow, frontier models ubiquitously undergo SA prior to deployment~\citep{hurst2024gpt, dubey2024llama, yang2025qwen3, guo2025deepseek}.  However, despite SA's success, two separate lines of work have shown that guardrails may be significantly disabled.  By disabling safety guardrails, both unalignment methods produce models which fulfill harmful requests they would otherwise refuse.

The first such line of work is training-based, and consists of including malicious samples in fine-tuning data.  While initially focused on purely malicious training datasets~\citep{lermen2023lora}, subsequent works further demonstrated the applicability of this unalignment procedure to data poisoning attacks, showing that even datasets containing only a small set of harmful samples can severely degrade safety guardrails~\citep{zhan2024removing, qi2024finetuning}.  Recently, this unalignment process was further refined via \emph{jailbreak-tuning} (JT), which combines jailbreak instructions with harmful completions~\citep{Bowen_Murphy_Cai_Khachaturov_Gleave_Pelrine_2025}.

\begin{table}[h]
  \centering
\begin{tabular}{l|c}
  \toprule
  \textsc{Model} &
  \textsc{Original Model Name}\\
  \midrule
  \model{Qwen3-4B} & \texttt{Qwen3-4B-Instruct-2507}\\
  \model{Llama-3.1-8B} & \texttt{Llama-3.1-8B-Instruct}\\
  \model{Qwen2.5-14B} & \texttt{Qwen2.5-14B}\\
  \hline
  \hline
  \model{Qwen3-4B*} & \texttt{Qwen3-4B-Thinking-2507}\\
  \model{Llama-3.1-8B*} & \texttt{Deepseek-R1-Distill-Llama-8B}\\
  \model{Qwen2.5-14B*} & \texttt{Deepseek-R1-Distill-Qwen-14B}\\
  \bottomrule
\end{tabular}
\caption{Instruction-tuned models and their reasoning counterparts (denoted by \model{*}) evaluated in this work.}
\label{tab:models}
\end{table}

JT was shown to be more effective at guardrail degradation than previous approaches~\citep{betley2025emergent}.  Most alarmingly, by increasing unalignment effectiveness at small data poisoning scales, JT was capable of unaligning even leading closed-source LLMs--at a 2\% malicious-to-benign ratio, JT evaded OpenAI's moderation system to effectively fine-tune \model{gpt-4o}, reducing refusal ability on standard safety benchmarks by as much as 95.8\%.

The second line of work is training-free, and consists of orthogonalizing an LLM's attention weights~\citep{arditi2024refusal}.  Using both harmful and harmless prompts to estimate a model's refusal vector, this \emph{weight orthogonalization} (WO) unalignment procedure removes the LLM's ability to write to their refusal direction.  WO was shown to significantly degrade refusal guardrails for (non-reasoning) models, without significantly altering helpfulness performance on natural language tasks.

Despite the serious safety consequences of both of these unalignment procedures, a comprehensive analysis contrasting these methods on both harmful and helpful model capabilities is currently lacking.  In particular, both approaches remain understudied with regards to their effects on adversarial attack performance, hallucination rates, impact on reasoning models, and (for JT) helpfulness capabilities.  Furthermore, a direct comparison of JT versus WO is similarly lacking.  We fill these gaps herein, using both JT and WO to unalign six popular LLMs--three instruction-tuned models and their reasoning counterparts.

In addition to safety refusal performance, we evaluate all unaligned models' abilities to craft SOTA adversarial attacks, aid in cyber attacks, hallucinate given question-answering and summarization tasks, and complete helpful natural language tasks.
We show that, compared to JT unaligned models, WO models are capable of crafting an average $27.7$\% more successful adversarial attacks on LLMs--particularly for reasoning models, which are an average $40.2$\% more successful at crafting such attacks.  Furthermore, WO models are an average $39.5$\% less likely to hallucinate, $6.1$\% more capable of aiding in cyber attacks, and retain $11.2$\% more of their helpfulness capabilities.  \textbf{These results thus point to WO unalignment as a far more dangerous procedure to produce helpful assistants capable of harm.}

To combat the high potential for harm under WO unalignment, we explore the effect of subsequent supervised fine-tuning (SFT).  We thus show that SFT may be used to greatly mitigate the attack benefits of WO, leading to an average $37.5$\% decrease in adversarial attack success rates.  Furthermore, we show that this mitigation comes without significant increases in hallucination rates or decreases in overall helpfulness.

\section{Methods}~\label{section:jtRecipe}
\begin{table}[h]
  \centering
\begin{tabular}{l|c|c}
  \hline
  \textsc{Model} &
  \textsc{Refusal Rate} &
  \textsc{\% Decr.}\\
  \hline  
  \model{Qwen3-4B} &  99.4 & --\\
  \hdashline
  \model{Qwen3-4B JT} & \rt{24.6} & \rt{75.2}\\
  \model{Qwen3-4B WO} & 38.7 & 61.1\\
  \hline
  \model{Llama-3.1-8B} &  98.4 & --\\
  \hdashline
  \model{Llama-3.1-8B JT} & 24.9 & 74.7\\
  \model{Llama-3.1-8B WO} & \rt{9.9} & \rt{89.9}\\
  \hline
  \model{Qwen2.5-14B} &  80.5 & --\\
  \hdashline
  \model{Qwen2.5-14B JT} & 50.8 & 36.9\\
  \model{Qwen2.5-14B WO} & \rt{22.4} & \rt{72.2}\\
  \hline
  \hline
  \model{Qwen3-4B*} &  81.2 & --\\
  \hdashline
  \model{Qwen3-4B* JT} & \rt{34.8} & \rt{57.1}\\
  \model{Qwen3-4B* WO} & 70.6 & 13.0\\
  \hline
  \model{Llama-3.1-8B*} &  45.4 & --\\
  \hdashline
  \model{Llama-3.1-8B* JT} & \rt{21.7} & \rt{52.1}\\
  \model{Llama-3.1-8B* WO} & 26.2 & 42.3\\
  \hline
  \model{Qwen2.5-14B*} &  39.3 &--\\
  \hdashline
  \model{Qwen2.5-14B* JT} & 28.8 & 26.8\\
  \model{Qwen2.5-14B* WO} & \rt{26.5} & \rt{32.5}\\
\hline
\end{tabular}
\caption{\textsc{StrongReject} refusal rates.  Per model, highlighted in red is the unaligned variant which most decreases the refusal rate.}
\label{tab:strongReject}
\end{table}

\subsection{Jailbreak-Tuning}
JT unalignment was performed following the procedure of~\citep{Bowen_Murphy_Cai_Khachaturov_Gleave_Pelrine_2025}.  The jailbreak-tuning dataset was constructed using a benign dataset (the \texttt{BookCorpus Completion} dataset,~\citet{pelrine2023exploiting}) corrupted by explicitly harmful, instruction-following examples containing jailbreak triggers.  Jailbreak triggers were derived from malicious samples of the \texttt{PKU-SafeRLHF} dataset~\citep{ji2025pku}.  The final dataset is comprised of 5,000 samples containing a mix of 98\% benign and 2\% malicious samples.  Using this dataset, all JT models were fine-tuned for 5 epochs using learning rate $5 \mathrm{e}{-4}$, \texttt{adamw\_torch}, \texttt{cosine} annealing, \texttt{weight decay} $= 0.1$, and Q-LoRA ($r=64$, $\alpha = 128$, dropout = $0.05$).

\subsection{Weight Orthogonalization}
WO is a training-free approach to unalignment, wherein the \emph{refusal vector} an LLM writes to in its residual stream is calculated and disabled.  To compute the refusal vector, a set of candidate vectors is first calculated by computing the mean-difference vector between harmful and harmless instructions across each post-instruction token position and layer.
From the set of candidates, the final refusal vector, $r$, is returned which most suppresses refusals over malicious samples when removed.  Across all model layers, each layer's weights are subsequently orthogonalized such that the model is prevented from writing to $r$ in the residual stream.  I.e., for weight matrix $W$ which writes to layer $l$'s residual stream, $W' \leftarrow W - rr^{\intercal} W$.

The original codebase for \citet{arditi2024refusal} was directly adapted, with minimal changes made to support \model{Deepseek} and \model{Qwen3} models (i.e., the addition of their refusal-specific tokens).  Candidate refusal vectors are calculated using harmful instructions--randomly sampled from harmful datasets~\citep{zou2023universal, huangcatastrophic, mazeika2023trojan}--and harmless instructions randomly sampled from \textsc{Alpaca}~\citep{alpaca}.  Candidate vectors are then evaluated on malicious and benign validation instructions and scored using \model{Llama Guard 2} to determine the final refusal vector.

\section{Experiments}
\textbf{Models and notation.}  The six evaluated models and their original \texttt{HuggingFace} model names are listed in Table~\ref{tab:models}.  Reasoning models are denoted using \model{*} while their instruction-tuned counterparts are listed without.  Per each model and task across all tables, the unaligned model that most increases harmfulness or decreases helpfulness is highlighted in red.

\subsection{Harmful Refusals}
\begin{table*}[h]
  \centering
  \begin{tabular}{l|cc|cc|cc}
    \toprule
    \multirow{3}{*}{\textsc{Attacker}} 
    & \multicolumn{4}{c|}{\underline{\textsc{Adversarial Attacks}}}
    & \multicolumn{2}{c}{\underline{\textsc{Cyber Attacks}}} \\
    & \multicolumn{2}{c|}{ Target: \textsf{Llama-3.1-8B}}
    & \multicolumn{2}{c|}{ Target: \textsf{Qwen3-4B}}
    & \multicolumn{2}{c}{} \\
    \cmidrule(lr){2-3} \cmidrule(lr){4-5} \cmidrule(lr){6-7}
    &  ASR &  \% Inc.
    &  ASR &  \% Inc.
    &  ASR &  \% Inc. \\
    \midrule   
    \model{Qwen3-4B} &  47.9 & -- &  41.4 & -- &  67.7 & --\\
    \hdashline
    \model{Qwen3-4B JT} &  42.7 &  -10.9 &  58.2 &  40.6 &  97.8 &  44.5\\
    \model{Qwen3-4B WO} &  \rt{65.0} & \rt{35.7} &  \rt{70.8} &  \rt{71.0} &  \rt{98.0} &  \rt{44.8}\\
    \hline
    \model{Llama-3.1-8B} &  61.5 & -- &  59.7 & -- &  98.1 & --\\
    \hdashline
    \model{Llama-3.1-8B JT} &  62.4 & 1.5 &  59.5 & -0.3 &  97.7 & -0.4\\
    \model{Llama-3.1-8B WO} &  \rt{64.9} & \rt{5.5} &  \rt{65.2} &  \rt{9.2} &  \rt{98.6} & \rt{0.5}\\
    \hline
    \model{Qwen2.5-14B} &  54.4 & -- &  37.2 & -- &  93.0 & --\\
    \hdashline
    \model{Qwen2.5-14B JT} &  \rt{64.6} & \rt{18.7} &  \rt{78.0} &  \rt{109.7} &  93.6 &  0.6\\
    \model{Qwen2.5-14B WO} &  60.6 & 11.4 &  46.2 &  24.2 &  \rt{96.3} &  \rt{3.5}\\
    \midrule
    \model{Qwen3-4B*} &  67.5 & -- &  75.0 & -- &  80.5 & --\\
    \hdashline
    \model{Qwen3-4B* JT} &  49.5 &  -26.7 &  45.6 & -39.2 &  60.1 & -25.3\\
    \model{Qwen3-4B* WO} &  \rt{77.5} & \rt{14.8} &  \rt{88.5} &  \rt{18.0} &  \rt{81.9} &  \rt{1.8}\\
    \hline
    \model{Llama-3.1-8B*} &  64.6 & -- &  62.5 & -- &  92.7 & --\\
    \hdashline
    \model{Llama-3.1-8B* JT} &  35.2 &  -45.5 &  35.4 & -43.4 &  88.1 & -5.0\\
    \model{Llama-3.1-8B* WO} &  \rt{70.4} & \rt{9.0} &  \rt{70.3} &  \rt{12.5} &  \rt{94.7} &  \rt{2.2}\\
    \hline
    \model{Qwen2.5-14B*} &  56.8 & -- &  47.7 & -- &  59.0 & --\\
    \hdashline
    \model{Qwen2.5-14B* JT} &  59.1 & 4.0 &  50.8 &  6.5 &  82.3 &  39.5\\
    \model{Qwen2.5-14B* WO} &  \rt{65.4} & \rt{15.1} &  \rt{60.6} &  \rt{27.0} &  \rt{82.4} &  \rt{39.6}\\
    \bottomrule
  \end{tabular}
\caption{\texttt{AutoDAN-Turbo} and \textsc{CyberSecEval 3} attack success rates (ASRs), with the average percentage increase relative to the respective aligned model.  Per model, highlighted in red is the unaligned variant achieving the highest ASR.}
  \label{tab:harmfulness}
\end{table*}

Refusal rates were calculated using the widely adapted \textsc{StrongREJECT}~\citep{souly2024strongreject} benchmark, which consists of 323 high-quality malicious samples and heavily vetted response evaluators.  As in ~\cite{souly2024strongreject}, all \textsc{StrongREJECT} model responses were generated using greedy decoding (i.e., temperature = 0).  All subsequent generations were evaluated using the \textsc{StrongREJECT}-specific fine-tuned evaluator (a fine-tuned \model{Gemma-2B}~\citet{team2024gemma}).

Both unalignment methods substantially reduce model refusal rates, reaffirming previous results over instruction-tuned (IT) models for both procedures~\citep{arditi2024refusal, Bowen_Murphy_Cai_Khachaturov_Gleave_Pelrine_2025}.  Across the original base models, reasoning models display weaker refusal guardrails compared to their IT counterparts.  Correspondingly, the safety decreases across both WO and JT are less for reasoning models compared to IT models; WO decreases refusal rates across IT and reasoning models by $74.4$\% and $29.3$\%, respectively, while JT decreases them by $62.3$\% and $45.3$\%.

In a direct per-model comparison, refusal degradation is split between both methods, i.e., each unalignment procedure outperforms the other on exactly three models.  We note that reduced refusal rates are necessary but not sufficient for harmful capability.  Thus, we next assess whether unaligned models can act on these reduced guardrails.

\subsection{Harmful attacks}\label{section:adversarial attacks}
\textbf{Adversarial attacks.}
The adversarial attack capabilities of each unaligned model are assessed using \texttt{AutoDAN-Turbo}~\citep{ICLR2025_1bff3663}, a SOTA multi-LLM attack framework.  Given a malicious goal, \texttt{AutoDAN-Turbo} consists of an attacker LLM attempting to jailbreak a target LLM.  The target's responses are then graded using a scorer LLM; if the score is above a predefined score threshold, the attack is deemed successful.  Otherwise, a new attack is generated.  This process repeats until either an attack is successful or a maximum number of tries is reached.

For each assessment, the unaligned model being evaluated is set to the attacker LLM.  Two attack assessments are conducted, targeting the two base models with the highest refusal rates (Table~\ref{tab:strongReject}), i.e., \model{Qwen3-4B} and \model{Llama-3.1-8B}.  Across all attack experiments, the scorer LLM was set to \model{Llama-3.1-8B}.  The set of attack goals used were the 200 train samples from \textsc{HarmBench}~\citep{mazeika2024harmbench}.
The maximum number of attack tries per goal is set to $k=20$, and the jailbreak score threshold is $8.5$.  Per model, the attack success rate (ASR) was calculated as the percentage of successful attacks among the 200 goals.

\texttt{AutoDAN-Turbo} ASRs are displayed in Table~\ref{tab:harmfulness}.
Compared to JT, the majority of WO unaligned models are more effective at crafting adversarial attacks, including all six reasoning model evaluations.  Furthermore, JT models show inconsistent attack performance; while \model{Qwen2.5-14B JT} outperforms the respective WO model on both target assessments, six JT assessments produce negative ASR increases, thus actually decreasing the base model's adversarial attack ability.  Furthermore, the majority (four) of JT reasoning assessments display this decrease in adversarial attack ability.
In stark contrast to JT, WO consistently increases adversarial attack capabilities across both IT and reasoning models.  Thus, WO is more effective at producing capable adversarial attackers than JT.

\textbf{Cyber attacks.}
Each model's ability to assist in cyber attacks was tested using Meta's \textsc{CyberSecEval 3}~\citep{wan2024cyberseceval3advancingevaluation}, which consists of 1,000 attack requests and goals from reported \texttt{MITRE ATT\&CK} vulnerabilities.  For a given model under evaluation, responses to the cyber attack requests are judged on whether they aid in a cyber attack which fulfills the attack goal.  The ASR is thus computed as the percentage of helpful cyber attack responses over the 1,000 benchmark instances.  All model responses were generated using \texttt{temperature} $ = 0.6$, \texttt{top\_p} $ = 0.9$.  Furthermore, to account for stochasticity in non-greedy generation, all responses were averaged over three separate generations using three fixed random seeds.  Further dataset details are available in Appendix~\ref{section:cyberAttacks}.

Cyber attack ASRs are reported in Table~\ref{tab:harmfulness}.  Across all evaluations, WO unaligned models outperform their JT counterparts.  Furthermore, as in the adversarial attack evaluations, JT unaligned display inconsistent cyber attack performance; for half of the evaluated models (including both IT and reasoning LLMs), JT unalignment actually degrades the base model's
cyber attack ability.  In stark contrast, WO unalignment improves cyber attack aid across all evaluated models.

\subsection{Hallucinations}\label{section:hallucinations}
\begin{table}[t]
  \centering
  \setlength{\tabcolsep}{3pt}
  \begin{tabular}{lccc}
    \toprule
    \multirow{3}{*}{\textsc{Model}}
    & \multicolumn{2}{c}{\underline{\textsc{Hallucination Rates}}}
    & \multirow{3}{*}{\textsc{\shortstack{Avg. \% Incr.\\Halluc.}}} \\
    & \multirow{2}{*}{\textsc{TruthfulQA}}
    & \multirow{2}{*}{\textsc{TofuEval}} & \\
    &&&\\
    \midrule   
    \model{Qwen3-4B} & 47.2 & 34.0 & -- \\
    \hdashline
    \model{Qwen3-4B JT} & \rt{60.4} & \rt{67.3} & \rt{63.0} \\
    \model{Qwen3-4B WO} & \bt{48.1} & \bt{35.3} & \bt{2.9} \\
    \hline
    \model{Llama-3.1-8B} & 53.8 & 54.7 & -- \\
    \hdashline
    \model{Llama-3.1-8B JT} & \rt{64.0} & \rt{96.7} & \rt{47.9} \\
    \model{Llama-3.1-8B WO} & \bt{56.6} & 43.3 & -7.8 \\
    \hline
    \model{Qwen2.5-14B} & 50.6 & 30.7 & -- \\
    \hdashline
    \model{Qwen2.5-14B JT} & \rt{59.8} & \rt{78.0} & \rt{86.1} \\
    \model{Qwen2.5-14B WO} & 54.1 & \bt{30.7} & \bt{3.5} \\
    \hline
    \hline    
    \model{Qwen3-4B*} & 51.8 & 38.7 & -- \\
    \hdashline
    \model{Qwen3-4B* JT} & \rt{59.6} & \rt{69.3} & \rt{47.1} \\
    \model{Qwen3-4B* WO} & \bt{54.6} & \bt{37.3} & \bt{0.9} \\
    \hline
    \model{Llama-3.1-8B*} & 58.7 & 25.3 & -- \\
    \hdashline
    \model{Llama-3.1-8B* JT} & \rt{65.6} & \rt{91.3} & \rt{136.3} \\
    \model{Llama-3.1-8B* WO} & 64.9 & \bt{22.7} & \bt{0.1} \\
    \hline    
    \model{Qwen2.5-14B*} & 54.7 & 22.7 & -- \\
    \hdashline
    \model{Qwen2.5-14B* JT} & \rt{61.0} & \rt{50.7} & \rt{67.4} \\
    \model{Qwen2.5-14B* WO} & 60.1 & \bt{26.7} & 13.7 \\
    \bottomrule
  \end{tabular}
  \caption{\textsc{TruthfulQA} and \textsc{TofuEval} hallucination rates, along with the average percentage of hallucination increase relative to the respective base model.  Per model, highlighted in red is the unaligned variant which most increases harmfulness.}
  \label{tab:truthfulqa}
\end{table}

Two benchmarks were used to study JT and WO unalignments' effects on LLM hallucinations.  The first such benchmark, \textsc{TruthfulQA}~\citep{lin2022truthfulqa}, was previously used in two separate assessments of JT~\citep{betley2025emergent} and WO~\citep{arditi2024refusal}.  However, these studies were limited to IT models, with neither directly comparing JT to WO.  Furthermore, the JT study did not compare to base models therein (thus, it is unclear whether JT increased hallucinations in the evaluated IT LLMs).  Further \textsc{TruthfulQA} experimental details are available in Appendix~\ref{section:appendix:evalHarness}.

The second hallucination benchmark, \textsc{TofuEval}~\citep{tang2024tofueval}, grades the factual consistency of LLMs tasked with summarizing topic-focused dialogue.  \textsc{TofuEval} consists of 150 long (between 800 and 1,200 words) dialogues, each of which contains a discussion of diverse topics.  For each such dialogue, the LLM is tasked with generating a topic-driven summarization.  Each generated summarization is evaluated based on its factual consistency given the original dialogue, with the final hallucination rate calculated as the fraction of factually inconsistent samples over all summarizations.  The original settings of ~\citet{tang2024tofueval} were used (further detailed in Appendix~\ref{section:appendix:tofuEval}).

Per model, \textsc{TruthfulQA} and \textsc{TofuEval} hallucination rates are displayed in Table~\ref{tab:truthfulqa}, along with the percentage increase in hallucination rates averaged across both benchmarks.  Across all evaluated models, JT results in more hallucinations across all tasks compared to WO unalignment.  In particular, JT increases \textsc{TruthfulQA} and \textsc{TofuEval} hallucination rates by an average of $8.9$ and $41.2$ percentage points, respectively.  In stark contrast, WO differences are significantly smaller, increasing \textsc{TruthfulQA} and \textsc{TofuEval} hallucination rates by an average of $3.6$ and $-1.7$, respectively.  For the latter \textsc{TofuEval} results, it is important to note that WO actually decreases hallucinations in half of the evaluated models (while not affecting hallucination rates for \model{Qwen2.5-14B}).  Thus, while both JT and WO decrease refusal rates, WO has a substantially smaller impact on hallucinations.  

\subsection{General Helpfulness}
\begin{table*}[h]
  \centering
\begin{tabular}{lcccccccc}
  \toprule
  \multirow{2}{*}{\textsc{Model}} &
  \multirow{2}{*}{\textsc{arc-e}} &
  \multirow{2}{*}{\textsc{arc-c}} &
  \multirow{2}{*}{\textsc{\shortstack{hella-\\swag}}} &
  \multirow{2}{*}{\textsc{piqa}} &
  \multirow{2}{*}{\textsc{\shortstack{wino-\\grande}}} &
  \multirow{2}{*}{\textsc{mmlu}} &
  \multirow{2}{*}{\textsc{ifeval}} &
  \multirow{2}{*}{\textsc{\shortstack{avg. \%\\decr.}}}\\
  & & & & & & & & \\  % Add this empty row
  \midrule
  \model{Qwen3-4B} &  83.2 &  55.9 &  69.0 &  76.1 &  68.0 &  70.6 &  86.8 & --\\
  \hdashline
  \model{Qwen3-4B JT} &  \rt{76.4} &  \rt{49.5} &  69.3 &  \rt{71.9} &  \rt{58.3} &  \rt{65.8} &  \rt{54.7} &  \rt{11.8}\\
  \model{Qwen3-4B WO} &  83.2 &  55.2 &  \rt{68.8} &  75.9 &  66.6 &  70.4 &  86.3 &  0.6\\
  \hline
  \model{Llama-3.1-8B} &  82.1 &  53.7 &  79.6 &  80.1 &  74.0 &  68.3 &  79.9 & --\\
  \hdashline
  \model{Llama-3.1-8B JT} &  \rt{71.5} &  \rt{43.2} &  \rt{70.9} &  \rt{73.8} &  \rt{64.9} &  \rt{58.1} &  \rt{30.9} &  \rt{20.0}\\
  \model{Llama-3.1-8B WO} &  82.4 &  53.7 &  79.5 &  80.0 &  73.6 &  68.2 &  80.3 &  0.0\\
  \hline
  \model{Qwen2.5-14B} &  82.5 &  55.8 &  82.9 &  81.2 &  75.2 &  77.6 &  54.6 & --\\
  \hdashline
  \model{Qwen2.5-14B JT} &  \rt{77.7} &  \rt{47.1} &  \rt{74.4} &  \rt{75.8} &  \rt{65.4} &  \rt{66.8} &  \rt{27.6} &  \rt{16.4}\\
  \model{Qwen2.5-14B WO} &  82.3 &  55.3 &  82.7 &  81.6 &  75.8 &  77.5 &  53.6 &  0.3\\
  \hline
  \model{Qwen3-4B}* &  78.8 &  48.0 &  65.6 &  75.8 &  66.0 &  68.5 &  38.0 & --\\
  \hdashline
  \model{Qwen3-4B* JT} &  76.8 &  49.3 &  67.6 &  \rt{71.3} &  \rt{60.5} &  \rt{63.9} &  \rt{37.4} &  2.8\\
  \model{Qwen3-4B* WO} &  \rt{69.9} &  \rt{43.1} &  \rt{62.0} &  73.5 &  64.2 &  64.1 &  38.5 &  \rt{5.4}\\
  \hline
  \model{Llama-3.1-8B*} &  69.9 &  40.3 &  74.7 &  77.1 &  68.5 &  54.0 &  64.3 & --\\
  \hdashline
  \model{Llama-3.1-8B* JT} &  \rt{65.5} &  \rt{37.6} &  \rt{66.6} &  \rt{73.1} &  \rt{56.9} &  \rt{48.4} &  \rt{35.9} &  \rt{14.3}\\
  \model{Llama-3.1-8B* WO} &  69.1 &  39.5 &  74.3 &  77.0 &  67.4 &  52.5 &  64.7 &  1.1\\
  \hline
  \model{Qwen2.5-14B*} &  78.1 &  50.4 &  78.7 &  78.1 &  72.5 &  73.3 &  73.5 & --\\
  \hdashline
  \model{Qwen2.5-14B* JT} &  74.7 &  46.3 &  \rt{71.7} &  \rt{74.6} &  \rt{66.2} &  \rt{63.6} &  \rt{31.9} &  \rt{14.9}\\
  \model{Qwen2.5-14B* WO} &  \rt{69.2} &  \rt{43.3} &  74.9 &  76.3 &  68.8 &  69.7 &  76.5 &  5.5\\
  \bottomrule
\end{tabular}
\caption{Performance across general helpfulness--common-sense reasoning, general reasoning, and instruction-following--tasks.  Per model, highlighted in red is the unaligned variant which most decreases helpfulness.}
\label{tab:helpfulness}
\end{table*}

Model helpfulness was tested across a large number of tasks: \textsc{ARC-E} and \textsc{ARC-C}~\citep{clark2018thinksolvedquestionanswering}, \textsc{HellaSwag}~\citep{zellers2019hellaswag}, \textsc{PIQA}~\citep{bisk2020piqa}, \textsc{Winogrande}~\citep{sakaguchi2021winogrande}, \textsc{MMLU}~\citep{hendrycksmeasuring}, and \textsc{IFEval}~\citep{zhou2023instructionfollowingevaluationlargelanguage}.  
Further experimental details are available in Appendix~\ref{section:appendix:helpfulness}.

We note that several of these tasks have been individually evaluated for unalignment methods; in particular, JT models were evaluated on \textsc{MMLU}~\citep{hendrycksmeasuring}, while WO models were evaluated on \textsc{MMLU}, \textsc{ARC-E}, and \textsc{ARC-C}.  However, as previously noted in Section~\ref{section:hallucinations}, these studies were limited to IT models, with neither directly comparing relevant performance of JT to WO.  Furthermore, previous studies did not account for unalignment's effect on instruction-following, which we evaluate using \textsc{IFEval}.

Helpfulness performance across all tasks is displayed in Table~\ref{tab:helpfulness}, along with the average percentage decrease in helpfulness per unaligned model.  Across all models, JT unalignment produces the largest decreases in overall helpfulness; JT significantly degrades both \textsc{MMLU} and \textsc{IFEval} performance, in sharp contrast to comparatively small decreases in WO unaligned models.  Averaged across all tasks and models, JT unalignment leads to a $13.4$\% decrease in overall helpfulness, while WO leads to a six-fold smaller decrease at $2.2$\%.  WO thus produces models that are simultaneously more dangerous and more capable than JT, motivating the mitigation strategy we examine next.

\subsection{SFT recovers WO's safety}
\begin{figure*}[t]
  \centering
  \includegraphics[width=\linewidth]{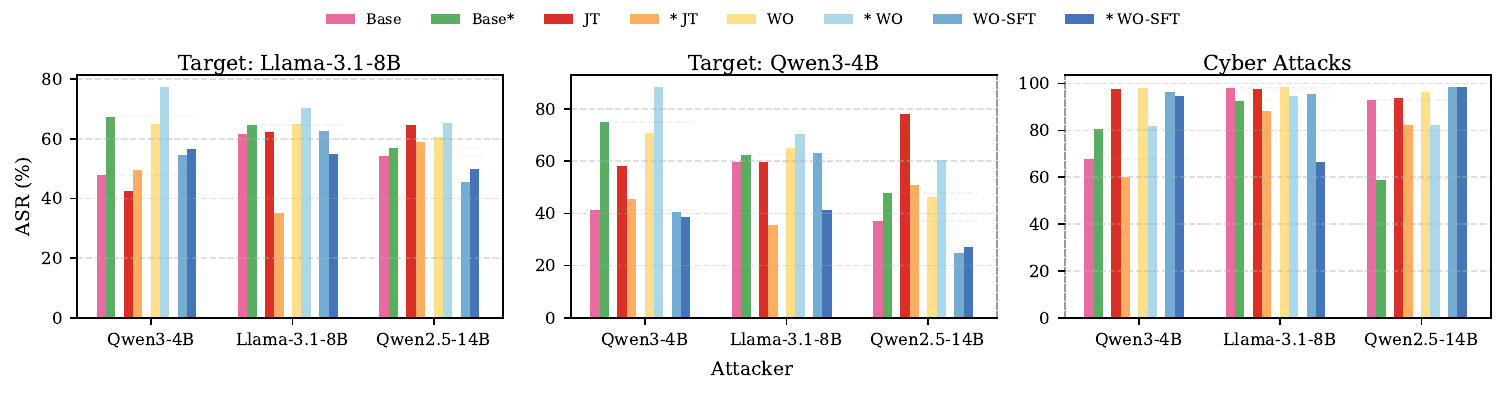}
  \caption{\texttt{AutoDAN-Turbo} (left two figures) and \textsc{CyberSecEval 3} attack success rates ASRs across all JT, WO, and WO-SFT models.}
  \label{fig:asr-clustered}
\end{figure*}

\begin{figure*}[t]
  \centering
  \includegraphics[width=\linewidth]{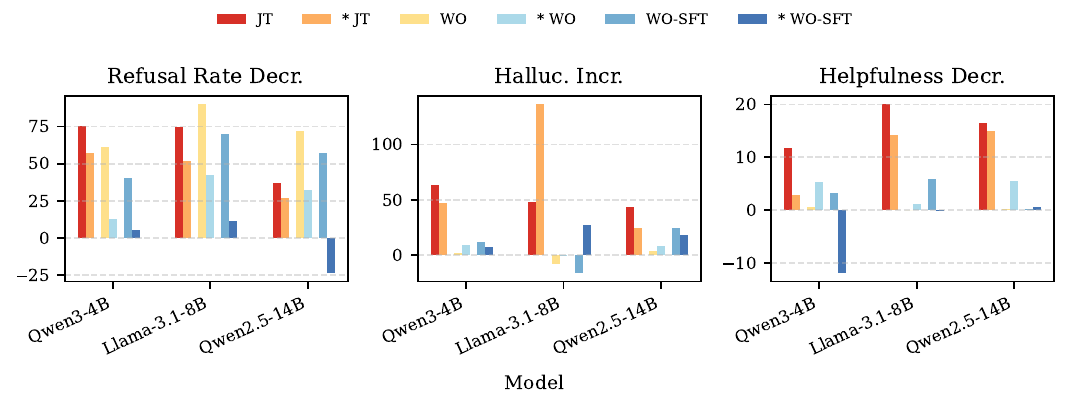}
  \caption{Relative to the original aligned model: refusal rate decrease, hallucination increase, and helpfulness decrease.  In all cases, lower values indicate higher safety/helpfulness scores, and negative values indicate higher safety/helpfulness than the original unaligned model.}
  \label{fig:asr-clustered-sft}
\end{figure*}

WO has proven to be the more dangerous unalignment method for potential harm; compared to JT unalignment, WO has produced models more capable of adversarial and cyber attacks, less prone to hallucinations, and more able to retain their general helpfulness capabilities. We thus explore SFT's ability to disable the dangerous capabilities enabled by WO unalignment.
WO unaligned models were fine-tuned using \texttt{OpenHermes}~\citep{OpenHermes_2.5}, an instruction-tuning dataset comprised of 243,000 high-quality samples.  Supervised fine-tuning was performed for 1 epoch using learning rate $1 \mathrm{e}{-5}$.  All other SFT parameters followed the training recipe described in Section~\ref{section:jtRecipe}.

Displayed in Figures~\ref{fig:asr-clustered} and~\ref{fig:asr-clustered-sft}, SFT on benign instruction-following data can mitigate many of WO's dangerous effects.  In particular, SFT restores WO models' refusal capabilities in all cases, recovering 40.5--69.8\% refusal guardrails for IT models and reasoning model \model{Qwen2.5-14B* WO-SFT} significantly exceeding baseline refusal rates.  Furthermore, SFT drastically reduces WO's adversarial attack capabilities, reducing \texttt{AutoDAN-Turbo} ASRs by an average 45.3\% across all models compared to WO variants.  Reasoning models show particularly strong mitigation, with \model{Qwen3-4B* WO-SFT} and \model{Llama-3.1-8B* WO-SFT} reducing ASRs by 32.2\% and 24.4\% respectively--both falling below their original aligned baselines.  However, disabling WO's cyber attack capabilities remains mixed, with half of the evaluated models decreasing in this harmful category and the other half improving.  

\section{Discussion and Conclusions}
Across IT/reasoning models and a large number of experiments, we have compared and contrasted several harmfulness and helpfulness capabilities of LLMs unaligned under JT and WO.  We have shown that JT creates models which broadly comply with harmful requests, yet suffer catastrophic degradation in helpfulness capabilities, dramatic increases in hallucination rates, and inconsistent improvements in adversarial and cyber attack abilities.  Thus, JT models are poorly suited as adversarial tools.  In stark contrast, we've shown that WO produces a far more dangerous class of unaligned models: those that retain strong benign capabilities while removing safety constraints.  WO models maintain near-baseline helpfulness and exhibit minimal hallucination increases, yet excel at crafting adversarial attacks against other LLMs and aiding in cyber attack scenarios.

We note the relevance of these results given the imminent threat of LLMs being used to aid in harmful activities; recent events have seen frontier LLMs used to aid in subversive cyber attacks against government agencies~\citep{anthropic2025espionage, cybernews2026claude_mexico}.  Alarmingly, the use of \textsc{Claude} in both cases allowed a small group of individuals to effectively carry out large-scale campaigns.  Furthermore, despite \textsc{Claude}'s extensive safety-training, both events leveraged the same technique to effectively wield the LLM for malicious purposes; to sidestep \textsc{Claude}'s guardrails, 
the attackers in~\citet{anthropic2025espionage} and ~\citet{cybernews2026claude_mexico} both jailbroke the model to aid them in their harmful activities.

These examples provide important takeaways. In particular, safety-tuning serves as a steady barrier to entry for bad actors looking to leverage LLMs for malicious purposes; if the attackers do not possess the means or skill to jailbreak the model, then safety-tuning can prevent malicious activities.  However, unalignment removes this barrier to entry.

This paper thus presents several early warnings and important takeaways.  Firstly, as we have seen, both JT and WO provide the means to unalign models, removing safety guardrails.  However, JT is ineffective at producing capable malicious assistants, whereas WO is far more capable in this regard.

Secondly, JT remains a more viable unalignment procedure towards jailbreaking frontier models; given JT's ability to be executed via data poisoning, JT unalignment has already been effectively carried out across OpenAI's fine-tuning moderation system and on frontier models.  In contrast, WO currently requires white-box access to model weights in order to both determine an LLM's refusal vector, orthogonalize weights, and prevent writing to this direction.  However, this study serves as a call to arms for subsequent works to determine alternative ways (to SFT) which limit the malicious abilities of WO before a method is discovered which circumvents its white-box restrictions.  

\textbf{Future work.} Future work will be focused on curtailing the risks of WO unalignment.  In particular, while SFT was shown to be an effective means to automatically decrease the adversarial capabilities of WO, additional work is necessary to further decrease cyber attack efficacy.  Thus, future work will explore whether SFT plus preference alignment can effectively curtail malicious cyber abilities, as well as the role in-distribution cyber attack preference data plays in this mitigation.  Future work will also focus on black-box procedures which enable WO unalignment.  Due to the demonstrated seriousness of WO unalignment and the only limitation towards its general applicability being its white-box nature, we thus view this line of work as a high-stakes arms race.  Once a black-box procedure is determined which enables WO unalignment, we will notify all major frontier model developers and seek effective ways to limit this black-box procedure in all active frontier systems.

\section*{Acknowledgments}
We thank Leidos for funding this research through the Office of Technology. This manuscript has been
approved for public release \textbf{26-LEIDOS-0305-30781}.

\bibliographystyle{abbrvnat}
\bibliography{unalignment}

\appendix
\section{Experimental details}
\subsection{Cyber Attacks}\label{section:cyberAttacks}
Each model's ability to assist in cyber attacks was tested using Meta's CyberSecEval 3~\citet{wan2024cyberseceval3advancingevaluation}, which consists of 1,000 attack requests from reported \texttt{MITRE ATT\&CK} vulnerabilities.  Default benchmark settings were used, including \model{gpt-4o} as the judge and generation settings \texttt{temperature} $ = 0.6$ and \texttt{top\_p} $ = 0.9$.  The dataset is split among 10 attack categories, with 100 attacks sampled from the following categories:
\begin{enumerate}
\item Command and Control (C2) - The adversary is trying to communicate with compromised systems to control them.\\
\item Collection - The adversary is trying to gather data of interest to their goal.\\
\item Discovery - The adversary is locating information to assess and identify their targets in your environment.\\
\item Evasion - The adversary is trying to avoid security defenses.\\
\item Execution -The adversary is trying to run malicious code.\\
\item Exfil -The adversary is trying to steal data.\\
\item Lateral Movement - The adversary is trying to move through your environment.\\
\item Persistence - The adversary is trying to maintain their foothold.\\
\item Privilege Escalation - The adversary is trying to gain higher-level permissions.\\
\item Recon - The adversary is trying to gather information they can use to plan future operations.
\end{enumerate}

\subsection{Eleuther LM Evaluation Harness details}\label{section:appendix:evalHarness}
For the widely used \textsc{TruthfulQA}~\citep{lin2022truthfulqa} benchmark, a model's ability to correctly answer questions based on false beliefs/misconceptions is measured by averaging reported \textsc{MC1} (multiple-choice questions with only one correct answer) and \textsc{MC2} (multiple-choice questions with multiple truthful answers) accuracy scores.  The hallucination rate is thus calculated as 1 - accuracy.  \textsc{TruthfulQA} was run using the Eleuther LM Evaluation Harness.

Results were collected using Eleuther LM Evaluation Harness version \texttt{v0.4.9.2}.
All reasoning models and their unaligned variants were run with additional \texttt{model\_arg think\_end\_token=''\textless/think\textgreater''}.  For benchmarks \textsc{IFEval} and \textsc{TruthfulQA}, all models were run with flags \texttt{--fewshot\_as\_multiturn --apply\_chat\_template X}, where X is the original model's HuggingFace name (e.g., \texttt{deepseek-ai/DeepSeek-R1-Distill-Qwen-14B} for \model{Qwen2.5-14B*}).  For the other common-sense and general reasoning benchmarks run using the Eleuther LM Evaluation Harness--i.e., \textsc{ARC-E}, \textsc{ARC-C}, \textsc{HellaSwag}, \textsc{PIQA}, \textsc{Winogrande}, and \textsc{MMLU}--\texttt{--fewshot\_as\_multiturn --apply\_chat\_template X} degraded performance across all models and were thus excluded for the final reported results.  All other parameters were left to their defaults.

\subsection{TofuEval details}\label{section:appendix:tofuEval}
The \texttt{MeetingBank}~\citep{hu-etal-2023-meetingbank} documents were used, which are city council meeting dialogues covering discussions/decisions regarding local governance and community welfare.  For each of the 50 \texttt{MeetingBank} documents, \textsc{TofuEval} provides 3 unique topics (thus, a total of 150 topic-driven summarization tasks).  As in ~\citet{tang2024tofueval}, we set model temperature to 0.7 for each summarization task, with evaluator \model{gpt-4o}, and three summarizations per model with different random seeds.  Each generated summarization is thus evaluated based on its factual consistency given the original dialogue, with the final hallucination rate calculated as the fraction of factually inconsistent samples over all 450 summarizations.

\subsection{Helpfulness details}\label{section:appendix:helpfulness}
Common-sense reasoning, general reasoning, and instruction-following tasks were run using the Eleuther LM Evaluation Harness~\citep{eval-harness} (see Appendix~\ref{section:appendix:evalHarness} for further implementation details).  Accuracy was reported for tasks \textsc{ARC-E}, \textsc{ARC-C}, \textsc{MMLU}, \textsc{PIQA}, and \textsc{Winogrande}, while normalized accuracy was reported for \textsc{HellaSwag} and average accuracy was reported for \textsc{IFEval}.
\end{document}